\documentstyle[12pt]{article}

\begin{document}
\vspace{0.5in}
\begin{center}
{\Large \bf On Connection between Coefficient Functions for
Deep-Inelastic and Annihilation Processes}
\vspace{0.5in}

{\large \bf G.T. Gabadadze}$^{a,b}$,~~{\large \bf A.L. Kataev}$^b$\\
\vspace{0.1in}

{\it Bogoliubov Laboratory of Theoretical Physics, Joint Institute for
\\ Nuclear Research, Dubna 141980, Russia $^{a}$}

{\it Institute for Nuclear Research of the Russian Academy of Sciences,
\\Moscow 117312, Russia $^b$}

\end{center}
\vspace{0.1in}
\begin{center}
{\bf Abstract}
\end{center}

It has been shown that the one-loop behavior of the axial anomaly,
occurring when the axial current is appropriately normalized, leads to
the cancellation of the corrections of type  $C_F^N{\bar \alpha}_s^N,~~
(N\geq 1) $ in the Crewther relation for the coefficient functions
of deep-inelastic and annihilation processes. The arguments in favour
of the overall factorization of the factor $\beta({\bar \alpha}_s)/
{\bar \alpha}_s$ in all orders of perturbation theory in this relation are
presented.

\vspace{0.3in}

In paper \cite {BK} the question on the status of the Crewther relation
\cite {CREW} in QCD has been investigated. In fact, using the update results
of the multiloop calculations the relation between the coefficient
functions for the deep-inelastic and annihilation processes has been
considered. The authors of ref. \cite {BK} have pointed out various
interesting properties of this relation. First of all, it has been
shown that the corrections of type $C_F{\bar \alpha}_s$,
$C_F^2{\bar \alpha}_s^2$ and $C_F^3{\bar \alpha}_s^3$ are cancelled
in the product of coefficient function from the Bjorken sum rule for
polarized deep-inelastic lepton-hadron scattering and the Adler function
for the two-point correlator of electromagnetic currents. It has also been
pointed out that the surviving corrections in the second and third
orders of perturbation theory are grouped yielding the two-loop
$\beta$--function. The result obtained in ref.  \cite {BK} has the form

$$
C_{Bj}({\bar a}_s)C_R ({\bar a}_s) =1+{\beta^{(2)} ({\bar a}_s)\over
{\bar a}_s}[K_1 C_F{\bar a}_s+(K_2 N_F+K_A C_A+K_F C_F)C_F{\bar
a}_s^2]+O({\bar a}_s^4),  \eqno (1)
$$
where ${\bar a}_s={\bar \alpha}_s(\mu^2=Q^2)/4\pi$, $N_F$ is the number
of flavors, $C_A$ and $C_F$ are the Casimir operators
(in QCD $C_A=3,~~C_F=4/3$),
$\beta^{(2)} ({\bar a}_s)=\beta_{1} {\bar a}_s^2+\beta_{2} {\bar
a}_s^3+O({\bar a}_s^4)$ is the QCD $\beta$--function in the two-loop
approximation. It is important to point out that this function does not
contain the terms of type $C_F^{N-1}{\bar \alpha}_s^N,~~(N\geq 2) $.
The numerical multipliers in eq. (1) are determined  as
$$ K_1=(-{21\over 2}
+12\zeta (3));~~K_2=({326\over 6}-{304\over 6}\zeta(3));~~
K_A=(-{629\over 2}+{884\over3}\zeta (3));
$$
$$ K_F=({397\over
6}+136\zeta(3)-240\zeta(5)).
$$
The coefficient function $C_{Bj}$  from the Bjorken
sum rule for polarized deep-inelastic lepton-hadron scattering is determined
through the following operator product expansion

$$
i\int T
V_{\alpha}(x)V_{\beta}(0)e^{ipx}dx|_{|p^2|\rightarrow\infty}\simeq
C_{Bj}({\bar a}_s) {\varepsilon_{\alpha\beta\rho\lambda}
p^{\rho}\over p^2} {1\over 12}A^{(3)\lambda}(0)+...
$$
Here $ V_{\alpha}$ denotes the electromagnetic current,  $A^{(3)\lambda}$
is the third component of the axial isotriplet (interpolating current for
$\pi$--meson). The expression for this coefficient function
is known in the two-loop \cite {CBJ2} and three-loop
\cite {CBJ3} approximations of perturbation theory. In the leading
order it has the form
$C_{Bj}({\bar a}_s)=1-3C_F{\bar
a}_s+ O({\bar a}_s^2)$.  The quantity $C_R$ from eq. (1) is just
the coefficient function for the branching ratio of $e^+e^-$-annihilation
into the hadrons. This coefficient function is also known in the
two-loop \cite {CR2} and three-loop \cite {CR3} approximations.
The leading order result has the form
$$
C_R ({\bar a}_s) ={D ({\bar a}_s) \over
N_c}=1+3C_F{\bar a}_s+ O({\bar a}_s^2),
$$
where the Adler function $D ({\bar a}_s )$ is defined as
$$
D ({\bar a}_s )=-12\pi^2q^2{d\over dq^2}\Pi(q^2),~~~
i\int \langle 0|T A^{(3)}_{\alpha}(x)A^{(3)}_{\beta}(0)|0\rangle
e^{iqx}dx= (g_{\alpha\beta}q^2- q_\alpha q_\beta)\Pi(q^2).
$$

The aim of the present investigation is to elucidate the reason of
cancellation of the $C_F{\bar a}_s$, $C_F^2{\bar a}_s^2$ and
 $C_F^3{\bar a}_s^3$ corrections in the Crewther relation and to generalize,
if possible, this low to the higher orders of perturbation theory.
As it will be demonstrated below, the observed cancellation is
intimately related to the specific structure of the anomalous triangle
and the Adler-Bardeen theorem \cite {AB}.

Let us consider the following three-point correlation function
$$
T_{\mu\alpha\beta}(p,q)
=\int\langle 0|TA^{(3)}_\mu(y)V_\alpha(x)V_\beta(0)|0\rangle
e^{ipx+iqy}dxdy
=\zeta_1(q^2,p^2 )\varepsilon_{\mu\alpha\beta\tau}p^\tau+
$$
$$
+\zeta_2(q^2,p^2 )(q_\alpha\varepsilon_{\mu\beta\rho\tau}p^\rho
q^\tau-q_\beta\varepsilon_{\mu\alpha\rho\tau}p^\rho
q^\tau)
 +\zeta_3(q^2,p^2 )(p_\alpha\varepsilon_{\mu\beta\rho\tau}p^\rho
q^\tau+p_\beta\varepsilon_{\mu\alpha\rho\tau}p^\rho q^\tau),
\eqno (2)
$$
where the expansion over the three independent tensor structures
is used (the kinematical condition $pq=0$ is also assumed,
for details see ref. \cite {GP}).

Following the ideology of ref.  \cite {CREW}, we consider the
operator product expansion for this correlator in the limit when
$|p^2|\rightarrow\infty$.  Now, using the relation for the various
tensor structures \cite {GP}, it is easy to derive that

$$
T_{\mu\alpha\beta}(p,q)\rightarrow {1\over 12}{1\over
p^2}C_{Bj}({\bar a}_s)
\Pi(q^2)(q_\alpha\varepsilon_{\mu\beta\rho\tau}p^\rho
q^\tau-q_\beta\varepsilon_{\mu\alpha\rho\tau}p^\rho
q^\tau),
$$
consequently
$$
\zeta_2(q^2,p^2 )|_{|p^2|\rightarrow\infty}\rightarrow {1\over 12}
{1\over p^2}C_{Bj}({\bar a}_s) \Pi(q^2).  \eqno (3) $$
On the other hand, requirement of the gauge invariance  leads one to
the Ward identity for the Green function under consideration.  In our case
the vector Ward identity takes the form \cite {GP}

$$
-\zeta_1(q^2,p^2 )=q^2\zeta_2(q^2,p^2 )+p^2\zeta_3(q^2,p^2 ).
$$
Differentiating this expression with respect of $q^2$ and taking
into account that the function $\zeta_1$ is just the nonrenormalizable
$c$--number (Adler-Bardeen theorem\cite {AB})
we get the following equation for two other invariant functions

$$
q^2{d\over dq^2}\zeta_2(q^2,p^2 )=-p^2 {d\over dq^2} \zeta_3(q^2,p^2
)- \zeta_2(q^2,p^2 ).   \eqno (4)
$$
It should be noticed here, that the statement on the one-loop behavior
of the axial anomaly has not strict sense within the perturbation theory.
On the language of operator relation the one-loop character is achieved
when the normalization of the axial current is strictly fixed in accordance
with the relation $ (\Lambda_\mu^5)^{Ren}=\gamma_5(\Lambda_\mu)^{Ren}$,
where $ (\Lambda_\mu^5)^{Ren}$ and $ (\Lambda_\mu)^{Ren}$ denote
the axial and vector vertex functions respectively. However, this condition
does not guarantee the absence of corrections on the language of Green
functions (in our case the absence of corrections to $\zeta_1$).
As it has been shown in ref. \cite {AnsJoh}, there are anomalous graphs
containing light-by-light subdiagrams which cause the renormalization
of the axial anomaly on the language of Green functions.
However, in our case when the axial current in (2) is the flavor
nonsinglet one, diagrams mentioned above renormalyze the quantity
$\zeta_1$ in the second order in the fine structure constant, but not in
${\bar a}_s^2$ order. Hence, neglecting the higher electromagnetic
corrections, we are able to postulate the one-loop character for
$\zeta_1$.

On the other hand, under the condition $|p^2|\rightarrow\infty$
and in accordance with the eq. (3) we have
$$
q^2{d\over dq^2}\zeta_2(q^2,p^2 )\rightarrow -{N_c\over (12\pi)^2}
{1\over p^2}C_{Bj} ({\bar a}_s) C_R({\bar a}_s).  \eqno (5)
$$
Let us now expand the expressions for the quantities $\zeta_2$
and  $\zeta_3$ in powers of $q^2/p^2$

 $$
\zeta_2(q^2,p^2)={1\over
p^2}\sum_{k=0}^{\infty}({q^2\over p^2})^k\zeta_2^k,~~~~~
\zeta_3(q^2,p^2)={1\over p^2}\sum_{n=0}^{\infty}({q^2\over
p^2})^n\zeta_3^n,
$$
here $\zeta_2^k$ and $\zeta_3^k$ are dimensionless coefficients.
Substituting these series into the eq. (4) one gets

$$
q^2{d\over dq^2}\zeta_2(q^2,p^2 )=-{1\over p^2}
\sum_{k=0}^{\infty}[(k+1)\zeta_3^{k+1}+\zeta_2^k]({q^2\over p^2})^k.
\eqno (6)
$$
Comparing now eq. (6) with the relation (3) we obtain the following
formulae for the product of $C_{Bj}$ and $C_R$

$$
{N_c\over (12\pi)^2} C_{Bj} ({\bar a}_s)
C_R({\bar a}_s)=\zeta_3^1+ \zeta_2^0.  \eqno (7)
$$
In the leading order of perturbation theory  $\zeta_3^1+ \zeta_2^0=N_c/
(12\pi)^2$. In so doing, we convinced ourselves that the one-loop
or a many-loop behavior for the product  $C_{Bj}C_R$ is connected
with the renormalizability or nonrenormalizability, respectively,
of the invariant functions $\zeta_2$ and $\zeta_3$.
On the other hand, it has been shown in  \cite
{Schrier}, that when the conformal invariance is exactly presented in
the theory, the general expression for the three-point correlator function
$T_{\mu\alpha\beta}$ has the form totally determined by its one-loop
counterpart $\Delta_{\mu\alpha\beta}$
$$
T_{\mu\alpha\beta}(p,q)=K({\bar a}_s) \Delta_{\mu\alpha\beta}(p,q),
$$
where $K({\bar a}_s)$ is the undefined quantity within the
approach of ref. \cite {Schrier}.
Another way of putting it is that in conformal-invariant theory
we have \cite {Schrier}
$$
\zeta_1^{exact}=K({\bar a}_s)\zeta_1^{one~loop},~~~\zeta_2^{exact}=
K({\bar a}_s)\zeta_2^{one~loop},~~~\zeta_3^{exact}=
K({\bar a}_s)\zeta_3^{one~loop}.  \eqno (8)
$$
However, it is well known that the renormalization
procedure violates the initial conformal invariance
of the massless QCD leading to the anomaly in the trace
of energy-momentum tensor \cite {CREW},\cite {CANOM}.
The expression for this anomaly \cite {CANOM} in its turn
indicates that the factor
$\beta({\bar a}_s)/({\bar a}_s)$ is the measure of violation
of conformal invariance within the framework of perturbation theory.
On this basis the relations (8) could be rewritten in QCD as
$$
\zeta_1^{exact}=K({\bar a}_s)\zeta_1^{one~loop},~~\zeta_2^{exact}=
[K({\bar a}_s)+{\beta({\bar a}_s)\over {\bar
a}_s}v_2(p^2,q^2, {\bar a}_s)]\zeta_2^{one~loop}, $$
$$
\zeta_3^{exact}=
[K({\bar a}_s)+{\beta({\bar a}_s)\over {\bar
a}_s}v_3(p^2,q^2,{\bar a}_s)]\zeta_3^{one~loop},
$$
where $v_2$ and  $v_3$ are dimensionless functions
satisfying to the Ward identity (4).
Arguing now, that in accordance with the Adler-Bardeen theorem
$\zeta_1^{exact}=\zeta_1^{one~loop}$, we obtain  $K({\bar a}_s)=1$.
Hence, the invariant functions $\zeta_2$ and $\zeta_3$
are renormalized in the higher orders of perturbation theory by
the multiplier containing the factor proportional to
 $\beta({\bar a}_s)/{\bar a}_s$ beyond the unity.  This fact leads to
the following expression for the product of
 $C_{Bj}$ and $C_R$
$$
C_{Bj}({\bar a}_s)C_R ({\bar a}_s) =1+{\beta({\bar a}_s)\over
{\bar a}_s}r({\bar a}_s),
$$
$r({\bar a}_s)$ being polynomial in powers of
${\bar a}_s$, which is not fixed in our approach.

In summary let us stress once again that in this work the reason of
cancellation of the $C_F^N{\bar \alpha}_s^N,~~(N\geq 1) $  type corrections
in the product of coefficient function from the Bjorken sum rule for
polarized deep-inelastic lepton-hadron scattering and the Adler function for
the two-point correlator of electromagnetic currents has been investigated.
It has been shown that the mentioned cancellation appears as a consequence of
the Adler-Bardeen theorem for the axial anomaly. It has also been
demonstrated that all surviving corrections are grouped producing the factor
proportional to the quantity $\beta({\bar a}_s)/{\bar a}_s$, which in its
turn is the measure of violation of conformal invariance in QCD.

The work is partly supported by the Russian Foundation for the
Fundamental Research, grant N94-02-04548a. The work of G.G. is
also supported by the International Soros Foundation, grant number N6J000.


\begin{thebibliography}{99}
\bibitem{BK}  D.J. Broadhurst, A.L. Kataev, Phys. Lett. B315(1993)179.
\bibitem{CREW}  R.J. Crewther, Phys. Rev. Lett. 28(1972)1421.
\bibitem{CBJ2} S.G. Gorishny, S.A. Larin, Phys. Lett. B172(1986)109;
\\E.B. Zijlstra, W.L. van Neerven, Phys. Lett. B297(1992)377.
\bibitem{CBJ3} S.A. Larin, J.A.M. Vermaseren, Phys. Lett. B259(1991)345.
\bibitem{CR2}  K.G. Chetyrkin, A.L. Kataev, F.V. Tkachov, Phys. Lett.
B85(1979)277;
\\M. Dine, J. Sapirstein, Phys. Rev. Lett. 43(1979)668;
\\W. Celmaster, R. Gonsalves, Phys. Rev. Lett. 44(1980)560.
\bibitem{CR3} S.G. Gorishny, A.L. Kataev, S.A. Larin, Pisma JETP
53(1991)121;
\\S.G. Gorishny, A.L. Kataev, S.A. Larin, Phys. Lett.
B259(1991)144;
\\ L.R. Surguladze, M.A. Samuel, Phys. Rev. Lett. 66(1991)560;
66(1991)2416 (erratum).
\bibitem{AB} S. Adler, W.
Bardeen, Phys. Rev. 182(1969)1517.
\bibitem{GP} G.T. Gabadadze, A.A. Pivovarov, Yad.Fiz. 56(1993)257;
\\Journ.Math.Phys. 35(1994)1045.
\bibitem{AnsJoh} A.A. Anselm, A.A. Johansen, JETP 96(1989)1181;
Yad. Fiz. 52(1990)882.
\bibitem{Schrier} E.J. Schrier, Phys. Rev. D3(1971)980.
\bibitem{CANOM} M.S. Chanowitz, J. Ellis, Phys. Lett. B40(1972)397;
Phys. Rev. D7(1973)2490; \\ J.C. Collins, A. Duncan, S.D. Joglekar,
Phys. Rev.  D16(1977)438; \\N.K. Nielsen, Nucl. Phys. B120(1977)212.
\end{thebibliography}
\end{document}